\documentclass[journal=jacsat,manuscript=article]{achemso}

\usepackage[utf8]{inputenc}
\usepackage{graphicx}
\usepackage[figurename=Fig.,labelfont=bf,labelsep=period]{caption}
\usepackage{xcolor}
\usepackage{amsmath}
\usepackage{amsfonts}
\title{Magneto-optics of the 2D iron-garnet nanocylinder array with localized and lattice modes}

\author{Polina E. Zimnyakova}
\affiliation[MIPT]
{Moscow Institute of Physics and Technology, National Research University, Dolgoprudny, Moscow, 141701 Russia}
\alsoaffiliation[Russian Quantum Center]
{Russian Quantum Center, Moscow, Russia}
\author{Daria O. Ignatyeva}
\email{ignatyeva@physics.msu.ru}
\affiliation[M.V. Lomonosov Moscow State University]
{Faculty of Physics, M.V. Lomonosov Moscow State University, Moscow, Russia}
\alsoaffiliation[Crimean Federal University]
{V.I. Vernadsky Crimean Federal University, Simferopol, Russia}
\alsoaffiliation[Russian Quantum Center]
{Russian Quantum Center, Moscow, Russia}
\author{Dolendra Karki}
\affiliation[Michigan Technological University]
{Physics Department, Michigan Technological University, Houghton, USA}
\author{Andrey A. Voronov}
\affiliation[MIPT]
{Moscow Institute of Physics and Technology, National Research University, Dolgoprudny, Moscow, 141701 Russia}
\alsoaffiliation[Russian Quantum Center]
{Russian Quantum Center, Moscow, Russia}
\author{Alexander N. Shaposhnikov}
\affiliation[Crimean Federal University]
{V.I. Vernadsky Crimean Federal University, Simferopol, Russia}
\author{Vladimir N. Berzhansky}
\affiliation[Crimean Federal University]
{V.I. Vernadsky Crimean Federal University, Simferopol, Russia}

\author{Miguel Levy}
\affiliation[Michigan Technological University]
{Physics Department, Michigan Technological University, Houghton, USA}
\author{Vladimir I. Belotelov}
\affiliation[M.V. Lomonosov Moscow State University]
{Faculty of Physics, M.V. Lomonosov Moscow State University, Moscow, Russia}
\alsoaffiliation[Crimean Federal University]
{V.I. Vernadsky Crimean Federal University, Simferopol, Russia}
\alsoaffiliation[Russian Quantum Center]
{Russian Quantum Center, Moscow, Russia}

\begin{document}

\begin{abstract}
We experimentally show the enhancement of the Faraday and transverse magneto-optical Kerr effects in the two-dimensional arrays of nanocylinders made of bismuth-substituted iron-garnet and supporting both localized and lattice modes. Simultaneous excitation of these modes makes it possible to increase the Faraday rotation by 3 times and TMOKE by an order of magnitude compared to the smooth magnetic film of the equal effective thickness. Both magneto-optical effects are enhanced in wide spectral and angular ranges making the nanocylinder array magnetic dielectric structures promising for applications with short and tightly-focused laser pulses. 
\end{abstract}

\section{Introduction}
    
Nowadays, the magneto-optical effects are widely used in different devices~\cite{maksymov2016magneto} such as routers~\cite{ho2018switchable}, optical isolators~\cite{karki2019broadband,karki2019toward,yan2020waveguide}, magneto-optical sensors~\cite{pourjamal2018hybrid,borovkova2020high,ignatyeva2016high,regatos2011suitable,inoue2011investigating,baryshev2013efficiency,qin2017ultrahigh}, modulators and magnetometers~\cite{ignatyeva2021vector,knyazev2018magnetoplasmonic,postava2004magneto,kuschel2011vectorial}. From the point of view of the device miniaturization, it is important to design the nanostructures with efficient magneto-optical interaction providing the enhancement of the magneto-optical effects compared to the same smooth films. 
    
First we focused on the enhancement of the magneto-optical Faraday effect which is the rotation of linear polarization of light passing through a material along an external magnetic field. The value of the rotation angle is directly proportional to the specific Faraday rotation of the material and the path traversed by light in the material. Since there is a dependence on the length of the path, miniaturization of devices inevitably leads to a decrease in the Faraday rotation value. Therefore, it is important to find ways to improve magneto-optical response of the materials. One of the straightforward ways to enhance the magneto-optical Faraday rotation in thin films is utilization of the photonic crystals with microcavity (defect) magnetic layers~\cite{mikhailova2018optimization,dong2020enhancing,steel2000photonic,lyubchanskii2006response,grishin2019waveguiding,Inoue_2006,ignatyeva2020bound,borovkova2020high,dadoenkova2017faraday,dyakov2019wide,pavlov2008enhancement}. Photonic crystals surrounding a magnetic layer act as a Bragg mirrors leading to the multiple re-reflection of light inside the magnetic layer like in the Fabry-Pérot cavity. Faraday polarization rotation constantly increases each loop the light travels inside the magnetic layer. Such amplification of the Faraday effect can be attributed to the increase in the effective path length of light through the material.

At the same time, it is possible to enhance magneto-optical effects via excitation of the optical resonances in the magnetic nanostructures, which can be attributed to the increase in the interaction time between light and a magnetic medium. Various types of nanostructures were shown to increase the magneto-optical effects~\cite{maccaferri2020nanoscale}. For example magneto-plasmonic crystals offer an opportunity to 
increase the Faraday and Kerr effects due to excitation of surface plasmon polaritons~\cite{caballero2015faraday,hamidi2015enhanced,kuzmin2016giant,uchida2009large,chekhov2018magnetoplasmonic,pomozov2020two,belotelov2006magnetooptics,borovkova2018tmoke,razdolski2013nonlocal}. The enhancement could be observed due to the excitation of plasmonic resonances in metallic particles or  nanoantennas~\cite{kharratian2020broadband,gabbani2021dielectric,zubritskaya2018magnetic,pourjamal2019tunable}, plasmonic nanocavities~\cite{lopez2020enhanced} and in the artificial metal-dielectric composites with hyperbolic dispersion~\cite{kolmychek2018magneto}. The possibility of the magneto-optical effects enhancement was also reported for structures that maintain simultaneously several kinds of optical modes~\cite{chin2013nonreciprocal,khramova2019resonances}. However, metals in such structures cause high absorption losses, broadening of observed resonances and decrease of the base (transmitted) signal. This problem can be solved by using all-dielectric resonant nanostructures~\cite{royer2020enhancement, bsawmaii2020longitudinal,voronov2020magneto,chernov2020all,ignatyeva2020all}. Recent studies~\cite{voronov2020magneto,ignatyeva2020all} show that besides the amplification of the Faraday rotation, one may obtain high magneto-optical intensity modulation in the transverse configuration of the external magnetic field applied to the structure. However, such amplification is observed in a very small angular and wavelength range due to the high Q-factor of the guided wave resonances.

Here we show the enhancement of the Faraday and transverse magneto-optical Kerr effect (TMOKE) in the two-dimensional arrays of cylinders made of bismuth-substituted iron-garnet that support both localized (Fabry–Pérot-like) and lattice (guided-like) modes. Simultaneous excitation of these modes makes it possible to increase the Faraday rotation by 3 times compared to the smooth magnetic film of the equal effective thickness. The one order increase of TMOKE is also observed in the structure in a wide angular range.

\section{Optical modes of the iron-garnet nanocylinder 2D array}

\begin{figure}
    \centering
    \includegraphics[width=0.5 \linewidth]{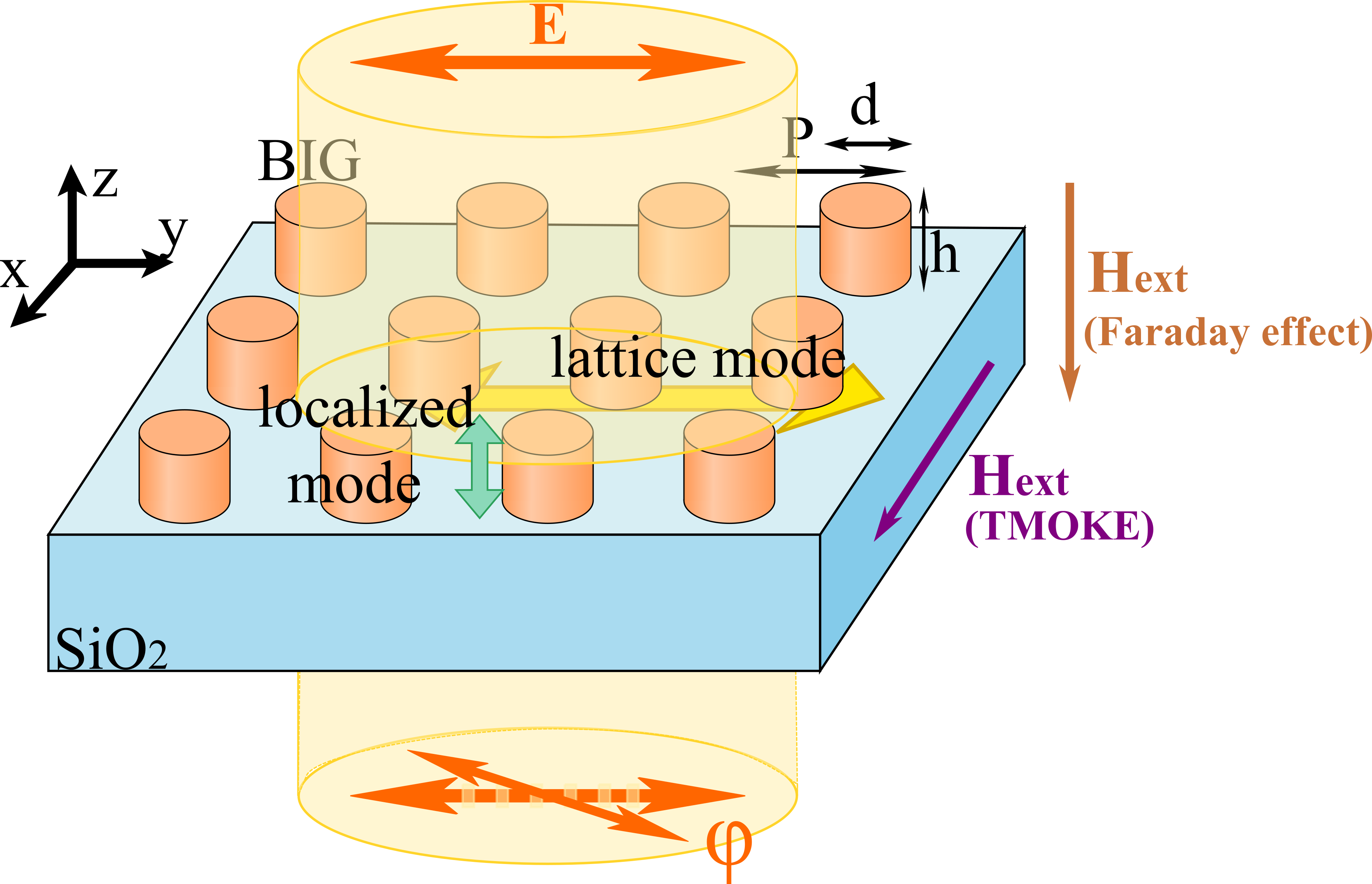}
    \caption{Schematic representation of the BIG nanocylinder 2D array and the excited localized and lattice modes.}
    \label{Fig.: scheme}
\end{figure}

The samples under research are two-dimensional arrays of cylinders etched in a bismuth substituted iron garnet (BIG) thin film (the thickness is 515 nm) of Bi$_{1.0}$Lu$_{0.5}$Gd$_{1.5}$Fe$_{4.2}$Al$_{0.8}$O$_{12}$/ Bi$_{2.8}$Y$_{0.2}$Fe$_{5.0}$O$_{12}$ deposited by magnetron sputtering on a SiO$_2$ substrate (Fig.~\ref{Fig.: scheme}). The nano-cylinders were patterned on a 550 nm thick spin-coated positive e-beam resist (ZEP-520A) by electron-beam exposure with a uniform dose of 140 µC/cm2 and under proximity effect correction (PEC) using a 100 KeV e-beam lithography system (VISTEC EBPG 5000+). A 30 nm-thick gold layer was also coated on top to avoid electrical charging of the dielectric garnet film during e-beam exposure. After which, the gold layer was first removed by wet etching in a gold etchant solution and then the resist was developed in an amyl acetate solution. The resist patterns were then transferred onto the BIG film by sputter-etching at a rate of 2.5 nm/minute with argon-ion beam. The temperature of the sample stage was maintained at 60C throughout the etching process to avoid hardening of the resist, which was then removed using resist remover N-methyl-2-pyrrolidine (NMP) by heating at 800C for about half an hour. The BIG cylinders with diameter $d$ ($d$=500~nm, 550~nm, 600~nm, 650~nm for the studied samples) having the same height $h=515$~nm were arranged in a square lattice with a period of $P$=900~nm. 

Such nanostructured material supports two kinds of the optical modes: the localized and lattice modes. The former could be understood in a frame of the waveguide theory as the modes of cylinder waveguides~\cite{adams1981introduction, snyder1986optical}. The waveguide with a circular cross-section supports various types of modes: TM and TE modes with axially-symmetric polarizations, and hybrid EH modes with the mixed polarizations. As in the present experiments the nanocylinders are illuminated by a rather wide linearly-polarized collimated light beam of $\sim 300~\mu$m diameter with a uniform distribution of the $\mathbf{E}$ vector at the scale of the nanocylinder, only modes with azimuth independent $\mathbf{E}$-field distribution could be excited. Actually, among waveguide modes only EH$_{1,m}$ modes (which correspond to the Linearly Polarized  LP$_{0,m}$ modes which are given by the characteristic equation $u[J_{m-1}(u)/J_m(u)]=-w[K_{m-1}(w)/K_m(w)]$ in the case of weak refractive index contrast\cite{gloge1971weakly}) have zero orbital number and azimuth independent spatial distribution of polarization vector inside the core. The field of such mode in the BIG cylinder has the form~\cite{adams1981introduction, snyder1986optical}: $E_x=J_{0}(2U_mr/d)\mathbf{e_x}$
where $\mathbf{e_j}$ is the unit polarization vector, $J_0$ is a Bessel function, $r=\sqrt{x^2+y^2}$ is the radial coordinate and $U_m=\frac{1}{2}k_0d\sqrt{n_\mathrm{BIG}^2 - n_m^2}$ is the dimensionless constant that depends on the mode refractive index $n_m$, $k_0$ is the wave vector of the incident light. 

\begin{figure}
    \centering
    (a)~~~~~~~~~~~~~~~~~~~~~~~~~~~~~~~~~~~~~~~~~~~~~~~~~~(b)\\
    \includegraphics[width=0.45\linewidth]{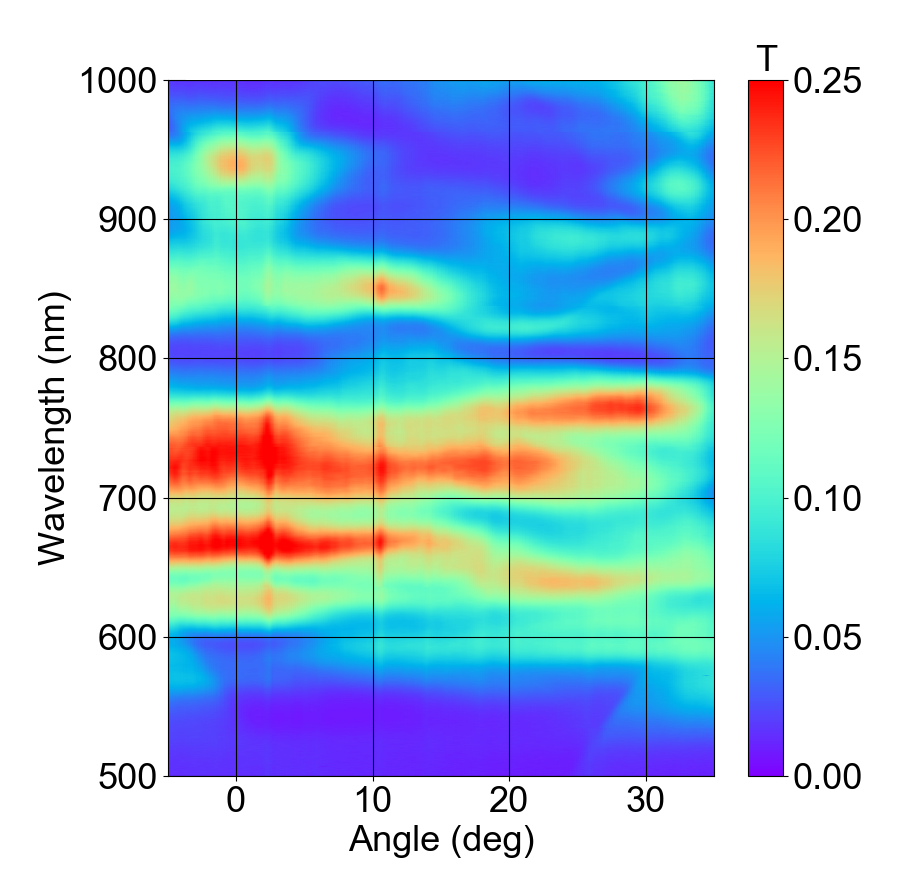}
    \includegraphics[width=0.45\linewidth]{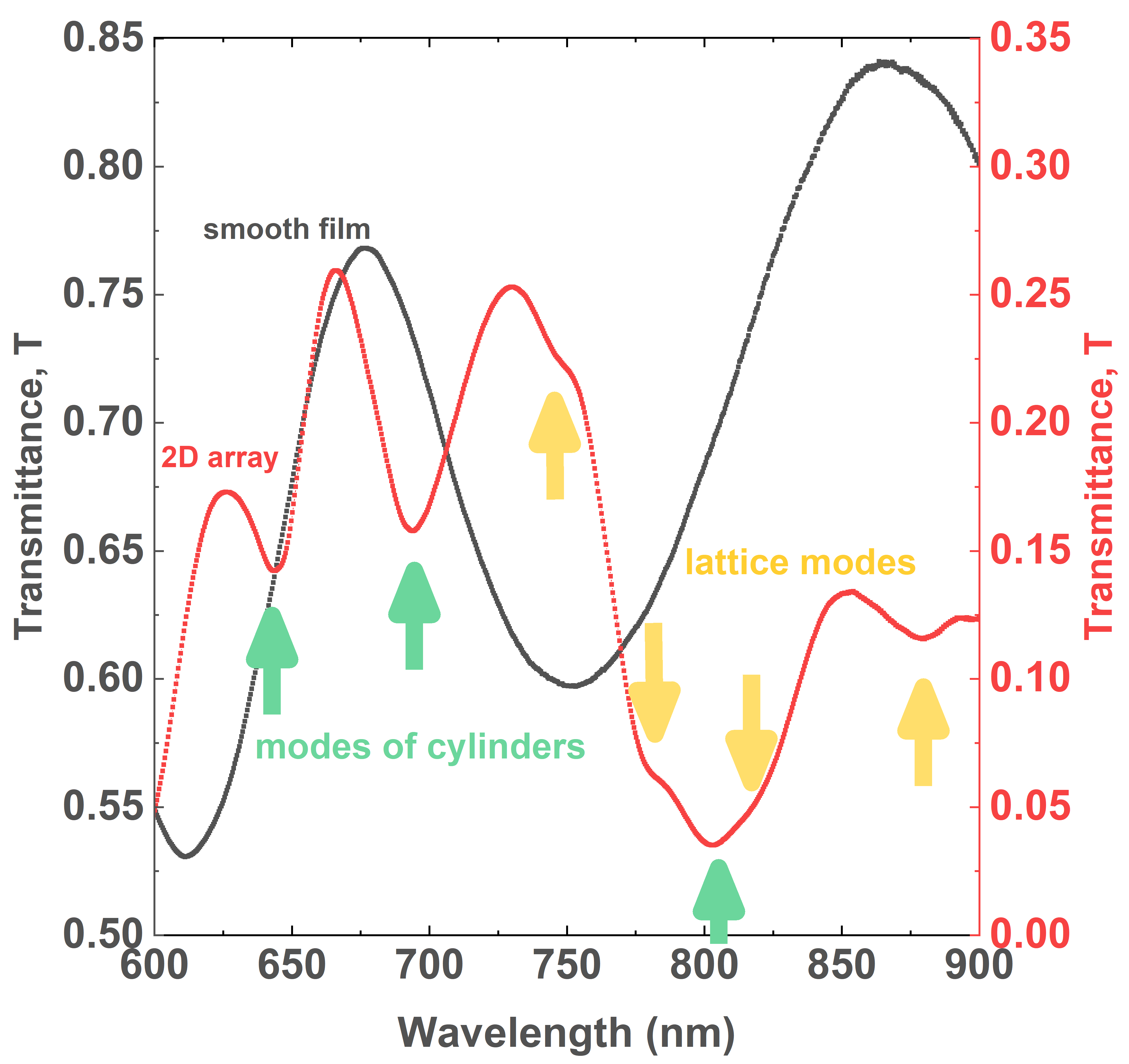}\\
        (c)~~~~~~~~~~~~~~~~~~~~~~~~~~~~~~~~~~~~~~~~~~~~~~~~~~(d)\\
    \includegraphics[width=0.9\linewidth]{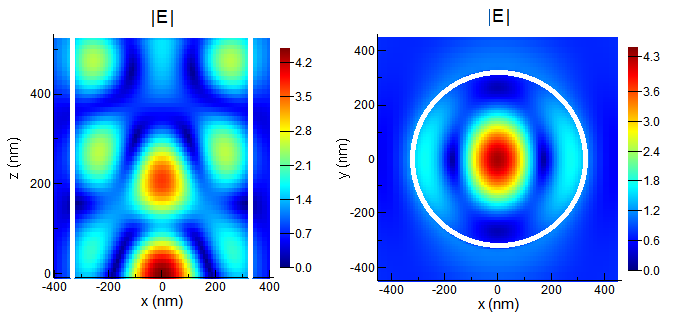}
    \caption{(a) False-color plot for the experimental dependence of the transmittance on the light incidence angle and wavelength for the structure with $d = 650$~nm; (b) Experimental transmittance spectra of the smooth film (black) and the array of cylinders (red) with $d = 650$~nm (the green (yellow) arrows indicate the dips corresponding to the excitation of the (lattice modes) eigenmodes). (c), (d) False-color plots for the spatial distributions of light electric field value $\mathrm{|E|}$ at the excitation of the localized mode at $\lambda=800~$nm in (c) $x-z$ and (d) $x-y$ planes.}
    \label{Fig.: Transmittance}
\end{figure}

Actually, we deal with a piece of the cylinder waveguide with the two facets neighboured by air ($n_{\mathrm{a}}=1$) and fused silica ($n_{\mathrm{SiO_2}}=1.45$), correspondingly. These facets act as the mirrors forming a 'vertical' Fabry–Pérot cavity (Fig.~\ref{Fig.: scheme}) and causing the minima and maxima in the transmittance spectra (see green arrows in Fig.~\ref{Fig.: Transmittance}b). The difference of the mode refractive indices $n_m$ for the modes of different orders leads to the differences in the resonant wavelengths. For example, as the numerical simulations show, the localized mode at $\lambda\sim800~$nm is the EH$_{1,2}$ mode (Fig.~\ref{Fig.: Transmittance}d) forming a standing wave (Fig.~\ref{Fig.: Transmittance}c) inside the BIG cylinder.

Obviously, the Q-factor of such a cavity is rather low, therefore the observed resonances are spectrally rather wide, 50-100~nm  in width, approximately. The positions of the observed resonances differ from the interference minima and maxima of the smooth BIG film (Fig.~\ref{Fig.: Transmittance}b), and are independent on the angle of incidence. These localized resonance spectral positions strongly depend on the diameter of the nanocylinder and experience a redshift if diameter increases.

Figure~\ref{Fig.: Transmittance}(a) shows that the angle-dependent lattice modes are also excited in the structure. These modes are formed due to the interaction of the leaking near-field radiation of the localized modes in the cylinders arranged periodically. Such modes can be treated as the guided modes propagating in the lateral direction in an effective planar waveguide with core formed by the nanopatterned BIG film ~\cite{voronov2020magneto,chernov2020all,ignatyeva2020all}. In a sense, these modes are similar to the lattice surface plasmons~\cite{rajeeva2018design}. 
In the 2D lattice, such guided modes can be excited under the following phase-matching condition:
\begin{equation}
 {\mathbf{k_\tau} + \mathbf{G_x}l_x + \mathbf{G_y}l_y = \boldsymbol{\beta}}
 \label{Eq. k_lattice}
\end{equation}
where $\mathbf{k_\tau}=k_0 \sin\theta \mathbf{e_\tau}$ is the tangential component of the incident light wavevector, $\theta$ is the angle of incidence, $l_x$ and $l_y$ are integers which correspond to the order of the lattice mode, $G_x=G_y=\frac{2\pi}{P}$ are the absolute values of the reciprocal lattice vectors, $P$ is the period of structure, $\lambda$ is the wavelength of the incidence light, $\boldsymbol{\beta}$ is the wave
vector of the lattice mode. As the period of the gratings is rather high ($P>\lambda$, $P>n_{\mathrm{SiO_2}}\lambda $) and there are several propagating diffraction orders generated in reflection and transmission, the efficiency of the lattice mode excitation is not very high. In Fig.~\ref{Fig.: Transmittance}a these modes reveal themselves intersecting the angle-independent resonances of the localized modes. In Fig.~\ref{Fig.: Transmittance}b lattice modes can be located as bents in the transmittance curve and are pointed with yellow arrows.

As it will be shown below, both of these modes are involved in the magneto-optical interaction of light with the 2D nanocylinder array. We will focus on the analysis of the localized mode EH$_{1,2}$ excited at $\lambda\sim800~$nm as it exists in all of the fabricated structures that makes it possible to track the evolution of the magneto-optical response with the variation of the nanocylinder diameter and appearance of the lattice modes in its vicinity.

\section{The Faraday effect with localized and lattice modes}

The magnetic film produces the magneto-optical Faraday rotation of the polarization of the transmitted light under the application of the external magnetic field parallel to the light wavevector. Such magnetization also influences the characteristics of the lattice and localized modes, in the different ways.

The localized modes of the nanocylinders have polarization degeneracy in the non-magnetic case, i.e. the two EH$_{1,m}$ modes of the same order $m$ and orthogonal polarizations $\mathbf{e}_x$ and $\mathbf{e}_y$ have the same refractive index $n_{1,m}$~\cite{snyder1986optical}. Actually, modes with circular polarization unit vectors $\mathbf{e}_x+i\mathbf{e}_y$ and $\mathbf{e}_x-i\mathbf{e}_y$ are the eigen modes of the system with the same refractive index $n_{1,m}$. Similar to the case of a smooth film, application of the external magnetic field along the cylindrical waveguide axes lifts this degeneracy so that the refractive indices of the both modes acquire additional magneto-optical  terms of different signs. This birefrigence between the two circularly polarized modes excited by the linearly-polarized incident light results in the Faraday rotation of its polarization. Notice that due to the complex dispersion of the nanocylinder modes such magneto-optical circular berifrigence differs from the one in a smooth film.

The lattice modes are also sensitive to the magnetization, however, the situation is more complicated. Actually, these modes have linear eigen polarization in the non-magnetic structure (for example, the TM-mode with $\{E_x,E_z,H_y\}$). External magnetic field applied normally to the structure changes these exact analytical solutions of Maxwell's equations and gives the different polarization of the eigen
modes. The eigenmode still has the same ($\{E_x,E_z,H_y\}$ for TM, for example) components but acquires linear in magnetization orthogonal (TE in the considered case $\{E_y,H_x,H_z\}$) components~\cite{kalish2014transformation}. The additional components are not associated with some other mode of the orthogonal polarization existing in the structure and appear just as a rigorous solution of Maxwell’s equations for the corresponding constitutive equations. 

It is important that the presence of the interface of the materials and the corresponding boundary conditions for the electromagnetic field also impose restrictions on the Faraday-like rotation of the polarization during the propagation along the surface~\cite{ignatyeva2012surface}. Thus quasi-TM and quasi-TE modes appear in the magnetized medium.  The refractive indices of the quasi-TM and quasi-TE modes remain the same as for the non-magnetic case. However, for rather thick magnetic films the dispersions of the quasi-TM and quasi-TE modes are very close to each other so that appearance of the magneto-optically induced orthogonal polarization components results in the energy swap between the quasi-TM and quasi-TE resulting in the polarization rotation~\cite{prokhorov1984optical}. This mechanism is less efficient as they the lattice modes in are rather shallow due to the high scattering during the propagation in lateral direction. Nevertheless, both types of the modes and both mechanisms are involved in the Faraday rotation observed in the structure as discussed below.

In order to compare the Faraday rotations in both cases correctly, we have to take into account the fact that the 2D arrays with different diameters of the nanocylinders have different specific amount of the magnetic BIG material per period. This amount is also different from the one in the smooth film of the same physical thickness. On the other hand, the specific Faraday rotation itself depends on the wavelength of the light thus it will inevitably differ for the resonances excited at different wavelengths. Therefore, to compare the nanostructure-induced enhancement of the Faraday rotation angles obtained in various structures more correctly, we calculate the value of relative enhancement as:
\begin{equation}
    \phi (\lambda) =\frac{4P^2}{\pi d^2} \frac{\Phi_{\mathrm{arr}}(\lambda)}{\Phi_\mathrm{film}(\lambda)}
\end{equation}
where $\Phi_{\mathrm{arr}}$ is the Faraday rotation angle measured in 2D array of cylinders, $\Phi_\mathrm{film}$ is the Faraday rotation angle measured in the smooth BIG film of the same thickness and composition, and the factor $\frac{4P^2}{\pi d^2}$ accounts for the variation of the relative amount of BIG material.

\begin{figure}
    \centering
    (a)~~~~~~~~~~~~~~~~~~~~~~~~~~~~~~~~~~~~~~~~~~~~~~~~~~(b)\\
    \includegraphics[width=0.45\linewidth]{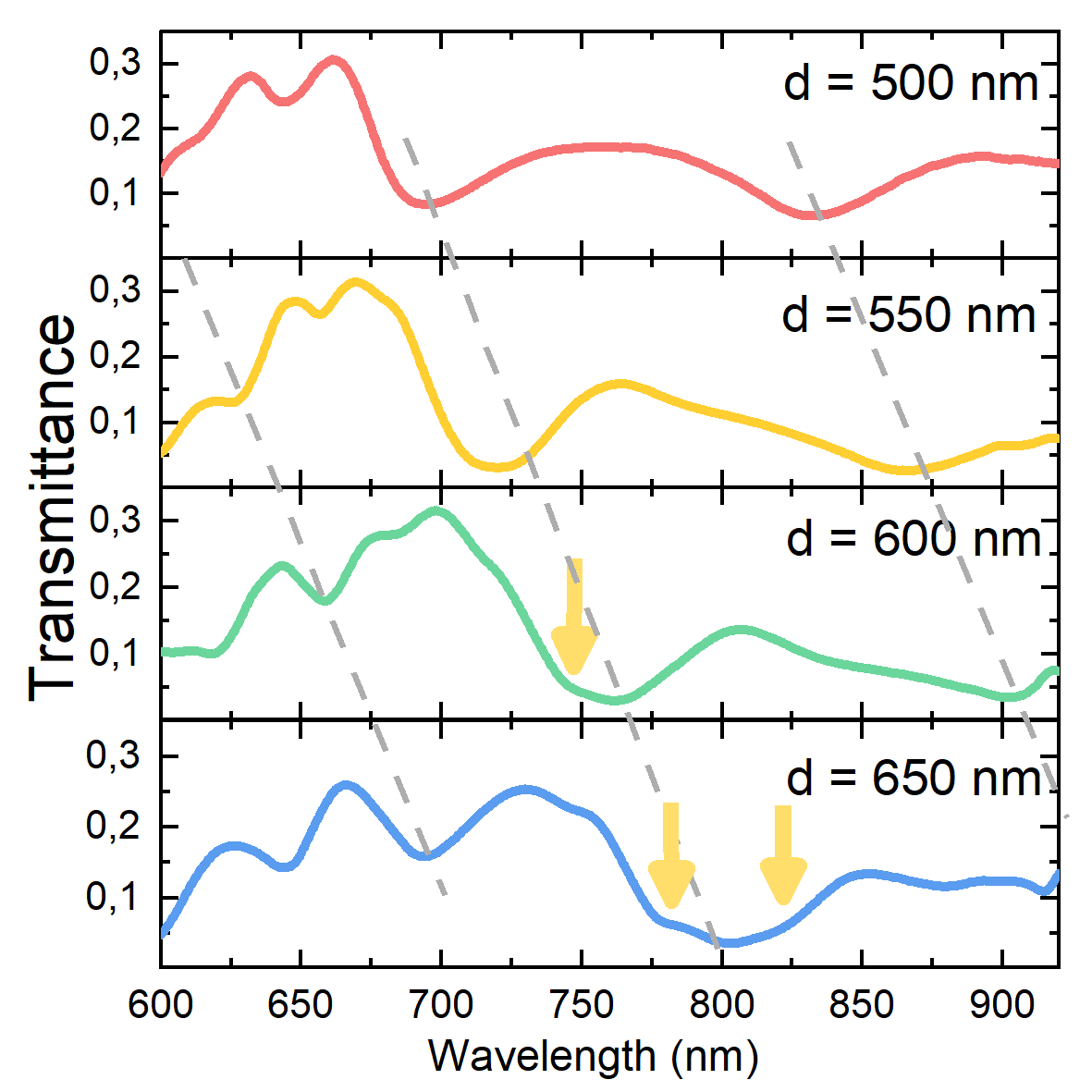}
    \includegraphics[width=0.45\linewidth]{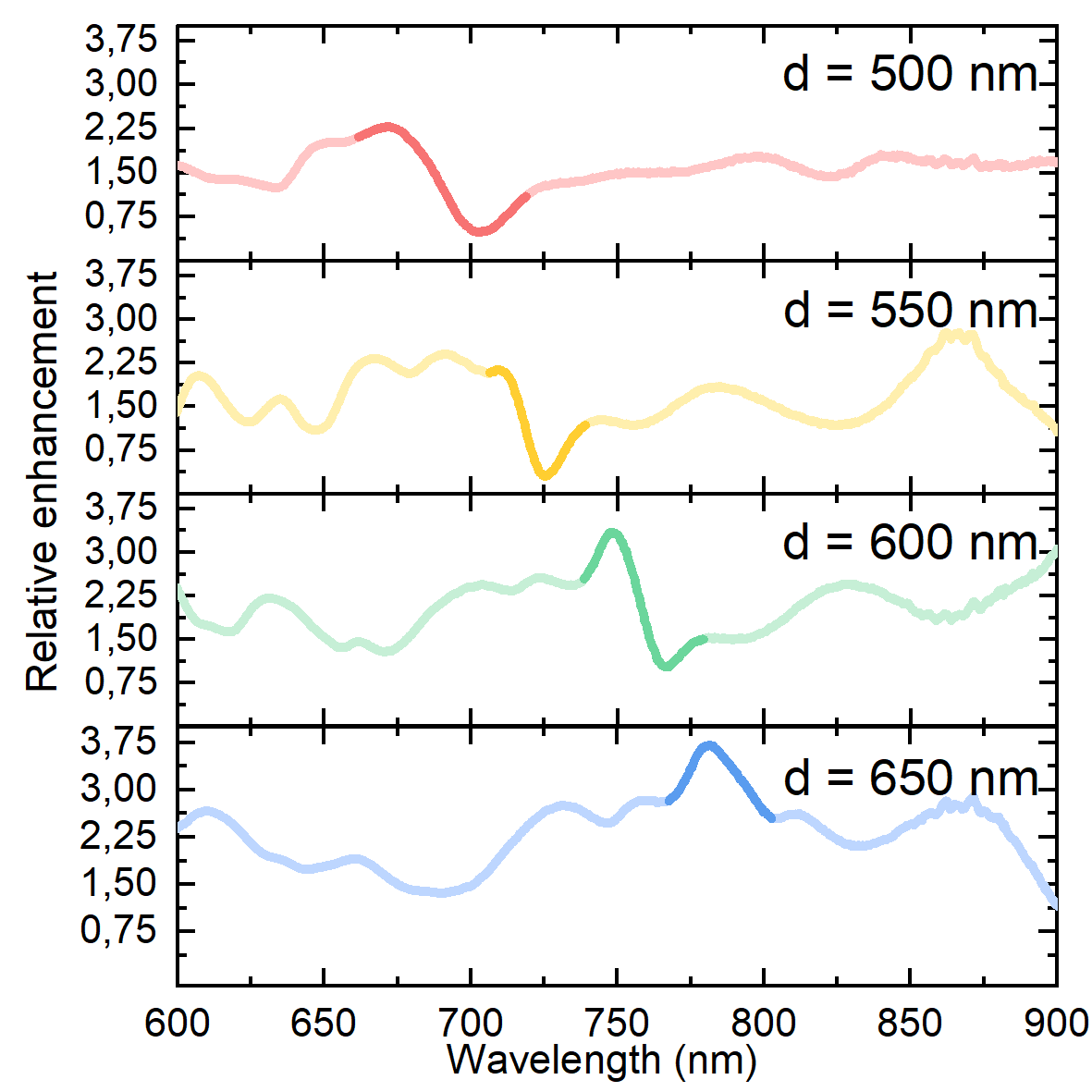}
    \label{fig:my_label}
    \caption{Relative enhancement of (a) transmittance and (b) the Faraday rotation in the nanocylinder arrays with different nanocylinder diameters (see the legends). In (b) the spectral range in the vicinity of EH$_{1,2}$ localized resonance is highlighted.}
\end{figure}

Figure 3(a) shows that for structures with diameters of cylinders $d=500~$nm and $d=550~$nm at the wavelengths in the vicinity of 700 nm, only the localized eigenmode of cylinders is excited. The relative enhancement of the Faraday effect in these structures is less than 2 times in comparison with a smooth BIG film. For structure with $d=600$ nm in the vicinity of the localized eigenmode excitation ($\lambda=770$ nm) the lattice mode is also excited so that the $\phi$ spectra acquire the Fano shape due to the superposition of these two resonances. Simultaneous excitation of the same eigenmode and two lattice modes in the structure with $d=650$ nm results in the interaction of the three magneto-optical resonances. This leads to the further change of the resonance shape which is more significant and makes the relative enhancement value larger up  to 3.75 times in comparison with that of a smooth iron garnet film. 

Therefore, the simultaneous excitation of the localized and lattice modes is shown to be responsible for nearly 4-times relative enhancement in Faraday rotation with respect to a smooth film of the same thickness.

\section{Transverse magneto-optical Kerr effect with localized and lattice modes}

The transverse magnetic field applied to the structure causes the magneto-optical Kerr effect that results into the modulation of the transmitted or reflected light intensity. This effect is below $10^{-4}$ in magnitude for the smooth iron-garnet films, however, it can be enhanced by the mode excitation.

The mechanism of the TMOKE amplification under the excitation of the lattice modes is the non-reciprocal magneto-optical variation of the mode propagation constant $\beta=\beta(H=0)+\Delta\beta(H)$ that changes the resonance conditions according to Eq.~\eqref{Eq. k_lattice} and causes a variation of the intensity of the transmitted and reflected light. The detailed description of this mechanism for all-dielectric gratings was presented elsewhere~\cite{voronov2020magneto}. Although it allows one to observe a strong enhancement of TMOKE, this enhancement is provided in a very narrow angular and frequency range in the vicinity of the lattice mode resonances. This limitation could be overcome if localized modes with angular-independent and wide resonances are excited simultaneously with the lattice ones.

\begin{figure}
    \centering
    (a)~~~~~~~~~~~~~~~~~~~~~~~~~~~~~~~~~~~~~~~~~~~~~~~~~~(b)\\
    \includegraphics[width=0.45\linewidth]{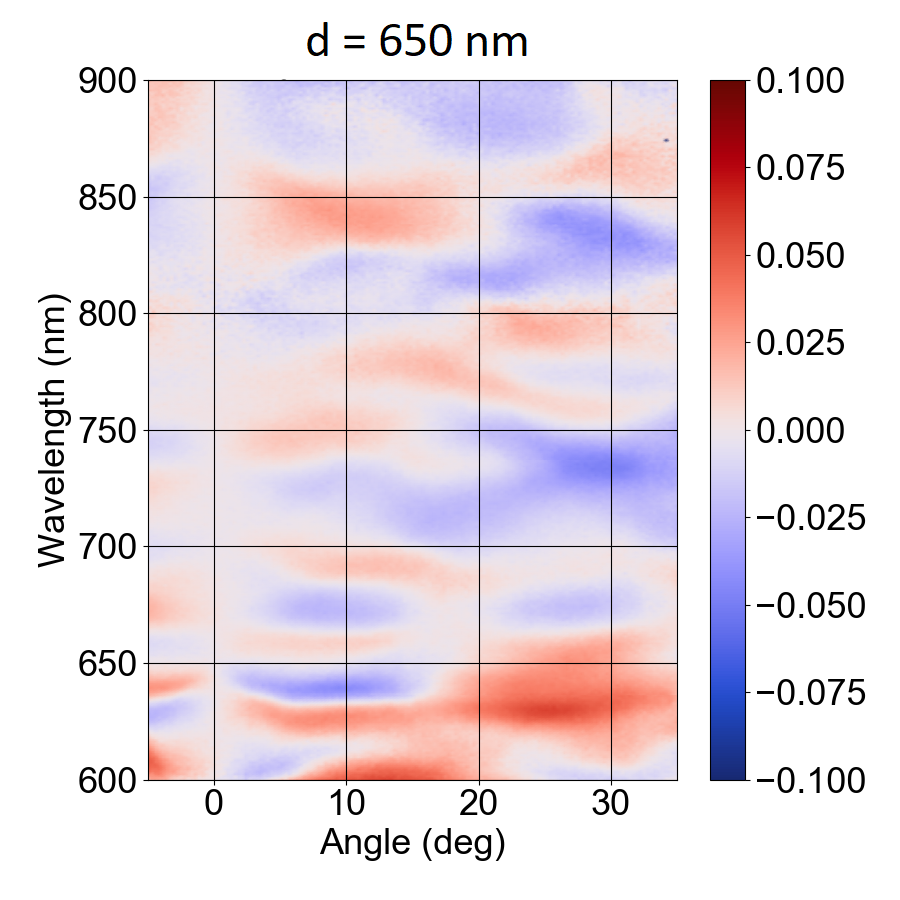}
    \includegraphics[width=0.54\linewidth]{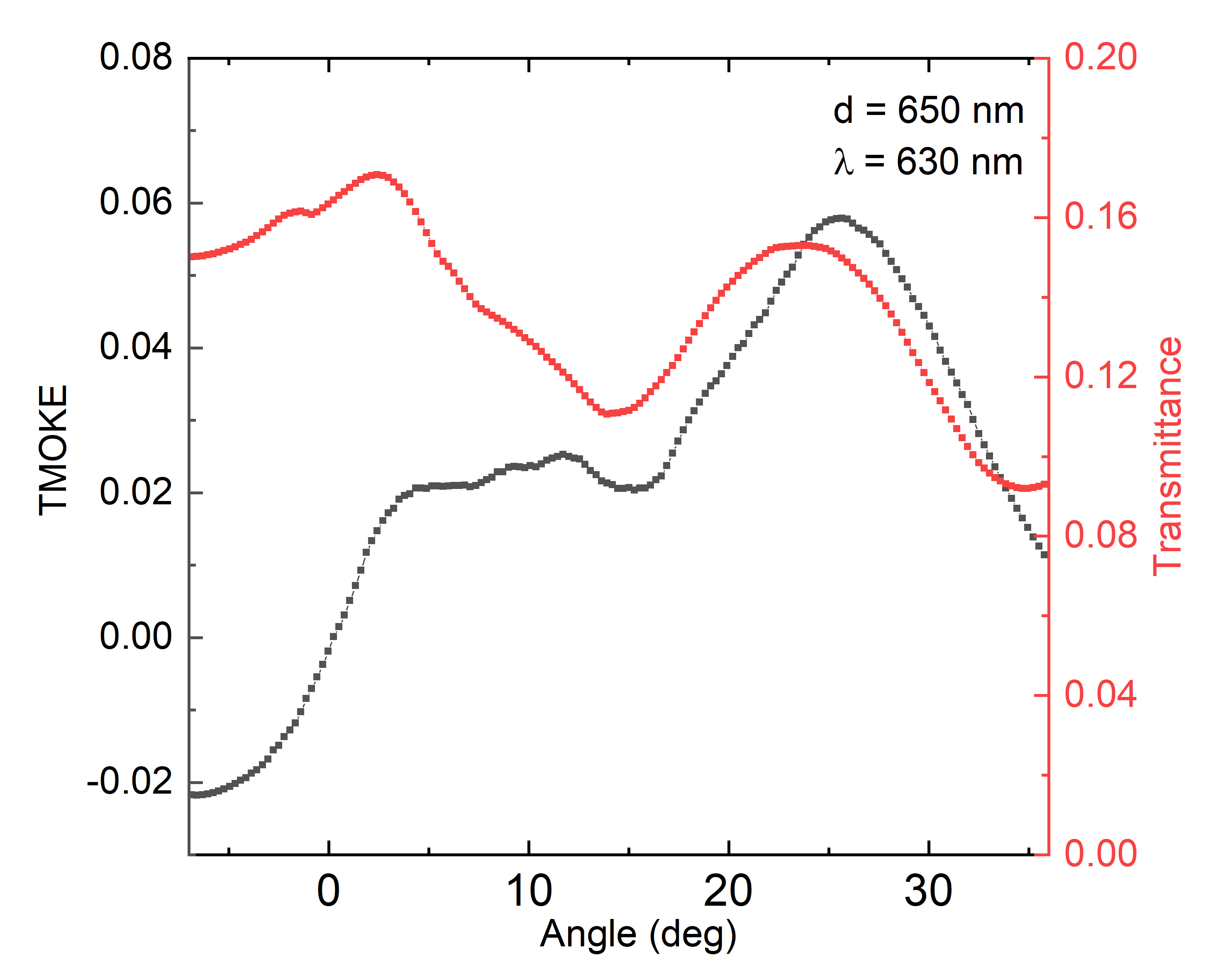}
    \caption{TMOKE in transmission for nanocylinder array with $d=650~$nm. (a) False-color plot of the dependence of TMOKE of the light with the incident angle and the wavelength. (b) Angular dependence of TMOKE and transmittance at the wavelength $\lambda=630~$nm. }
    \label{fig:TMOKE}
\end{figure}

Figure~\ref{fig:TMOKE} shows the TMOKE spectra for the nanocylinder array with $d=650~$nm. One may see that the TMOKE is enhanced in the whole measurement range and the resonances are broad and have large angular width. The most interesting is the TMOKE enhancement up to $5\%$ for $\lambda=630~nm$, and exceeds $2\%$ for the angles of incidence from 5 to 35 deg. It opens up new possibilities for the device miniaturization since such wide spectra allows one for the utilization of tightly-focused light for the efficient magneto-optical modulation.   

\section{Conclusion}

The enhancement of the Faraday and transverse magneto-optical Kerr effects in the two-dimensional arrays of nano-cylinders made of bismuth-substituted iron-garnet were shown experimentally. The feature of the considered structure is the co-existence of both localized (Fabry–Pérot-like) and lattice (guided-like)modes. Simultaneous excitation of these modes makes it possible to increase the Faraday rotation by 3 times compared to that in the smooth magnetic film of the equal effective thickness. The one order of magnitude increase in TMOKE is also observed in the structure. It is important that both the Faraday rotation and TMOKE are enhanced in wide spectral and angular ranges. This makes the considered structure prospective for magneto-optical applications with tightly-focused femtosecond laser pulses that usually have rather broad frequency and angular spectra. 

\section{Funding}
The study of the Faraday effect was financially supported by the Ministry of Science and Higher Education of the Russian Federation, Megagrant project No. 075-15-2019-1934, and the study of transverse magneto-optical Kerr effect was financially supported by RSF 21-72-10020.

\bibliography{references}

\providecommand{\latin}[1]{#1}
\makeatletter
\providecommand{\doi}
  {\begingroup\let\do\@makeother\dospecials
  \catcode`\{=1 \catcode`\}=2 \doi@aux}
\providecommand{\doi@aux}[1]{\endgroup\texttt{#1}}
\makeatother
\providecommand*\mcitethebibliography{\thebibliography}
\csname @ifundefined\endcsname{endmcitethebibliography}
  {\let\endmcitethebibliography\endthebibliography}{}
\begin{mcitethebibliography}{57}
\providecommand*\natexlab[1]{#1}
\providecommand*\mciteSetBstSublistMode[1]{}
\providecommand*\mciteSetBstMaxWidthForm[2]{}
\providecommand*\mciteBstWouldAddEndPuncttrue
  {\def\EndOfBibitem{\unskip.}}
\providecommand*\mciteBstWouldAddEndPunctfalse
  {\let\EndOfBibitem\relax}
\providecommand*\mciteSetBstMidEndSepPunct[3]{}
\providecommand*\mciteSetBstSublistLabelBeginEnd[3]{}
\providecommand*\EndOfBibitem{}
\mciteSetBstSublistMode{f}
\mciteSetBstMaxWidthForm{subitem}{(\alph{mcitesubitemcount})}
\mciteSetBstSublistLabelBeginEnd
  {\mcitemaxwidthsubitemform\space}
  {\relax}
  {\relax}

\bibitem[Maksymov(2016)]{maksymov2016magneto}
Maksymov,~I.~S. Magneto-plasmonic nanoantennas: basics and applications.
  \emph{Reviews in Physics} \textbf{2016}, \emph{1}, 36--51\relax
\mciteBstWouldAddEndPuncttrue
\mciteSetBstMidEndSepPunct{\mcitedefaultmidpunct}
{\mcitedefaultendpunct}{\mcitedefaultseppunct}\relax
\EndOfBibitem
\bibitem[Ho \latin{et~al.}(2018)Ho, Im, Pae, Ri, Han, and
  Herrmann]{ho2018switchable}
Ho,~K.-S.; Im,~S.-J.; Pae,~J.-S.; Ri,~C.-S.; Han,~Y.-H.; Herrmann,~J.
  Switchable plasmonic routers controlled by external magnetic fields by using
  magneto-plasmonic waveguides. \emph{Scientific reports} \textbf{2018},
  \emph{8}, 1--8\relax
\mciteBstWouldAddEndPuncttrue
\mciteSetBstMidEndSepPunct{\mcitedefaultmidpunct}
{\mcitedefaultendpunct}{\mcitedefaultseppunct}\relax
\EndOfBibitem
\bibitem[Karki \latin{et~al.}(2019)Karki, Stenger, Pollick, and
  Levy]{karki2019broadband}
Karki,~D.; Stenger,~V.; Pollick,~A.; Levy,~M. Broadband bias-magnet-free
  on-chip optical isolators with integrated thin film polarizers. \emph{Journal
  of Lightwave Technology} \textbf{2019}, \emph{38}, 827--833\relax
\mciteBstWouldAddEndPuncttrue
\mciteSetBstMidEndSepPunct{\mcitedefaultmidpunct}
{\mcitedefaultendpunct}{\mcitedefaultseppunct}\relax
\EndOfBibitem
\bibitem[Karki \latin{et~al.}(2019)Karki, El-Ganainy, and
  Levy]{karki2019toward}
Karki,~D.; El-Ganainy,~R.; Levy,~M. Toward high-performing topological
  edge-state optical isolators. \emph{Physical Review Applied} \textbf{2019},
  \emph{11}, 034045\relax
\mciteBstWouldAddEndPuncttrue
\mciteSetBstMidEndSepPunct{\mcitedefaultmidpunct}
{\mcitedefaultendpunct}{\mcitedefaultseppunct}\relax
\EndOfBibitem
\bibitem[Yan \latin{et~al.}(2020)Yan, Yang, Liu, Zhang, Xia, Kang, Yang, Qin,
  Deng, and Bi]{yan2020waveguide}
Yan,~W.; Yang,~Y.; Liu,~S.; Zhang,~Y.; Xia,~S.; Kang,~T.; Yang,~W.; Qin,~J.;
  Deng,~L.; Bi,~L. Waveguide-integrated high-performance magneto-optical
  isolators and circulators on silicon nitride platforms. \emph{Optica}
  \textbf{2020}, \emph{7}, 1555--1562\relax
\mciteBstWouldAddEndPuncttrue
\mciteSetBstMidEndSepPunct{\mcitedefaultmidpunct}
{\mcitedefaultendpunct}{\mcitedefaultseppunct}\relax
\EndOfBibitem
\bibitem[Pourjamal \latin{et~al.}(2018)Pourjamal, Kataja, Maccaferri,
  Vavassori, and Van~Dijken]{pourjamal2018hybrid}
Pourjamal,~S.; Kataja,~M.; Maccaferri,~N.; Vavassori,~P.; Van~Dijken,~S. Hybrid
  Ni/SiO2/Au dimer arrays for high-resolution refractive index sensing.
  \emph{Nanophotonics} \textbf{2018}, \emph{7}, 905--912\relax
\mciteBstWouldAddEndPuncttrue
\mciteSetBstMidEndSepPunct{\mcitedefaultmidpunct}
{\mcitedefaultendpunct}{\mcitedefaultseppunct}\relax
\EndOfBibitem
\bibitem[Borovkova \latin{et~al.}(2020)Borovkova, Ignatyeva, Sekatskii,
  Karabchevsky, and Belotelov]{borovkova2020high}
Borovkova,~O.; Ignatyeva,~D.; Sekatskii,~S.; Karabchevsky,~A.; Belotelov,~V.
  High-Q surface electromagnetic wave resonance excitation in magnetophotonic
  crystals for supersensitive detection of weak light absorption in the
  near-infrared. \emph{Photonics Research} \textbf{2020}, \emph{8},
  57--64\relax
\mciteBstWouldAddEndPuncttrue
\mciteSetBstMidEndSepPunct{\mcitedefaultmidpunct}
{\mcitedefaultendpunct}{\mcitedefaultseppunct}\relax
\EndOfBibitem
\bibitem[Ignatyeva \latin{et~al.}(2016)Ignatyeva, Kapralov, Knyazev, Sekatskii,
  Dietler, Nur-E-Alam, Vasiliev, Alameh, and Belotelov]{ignatyeva2016high}
Ignatyeva,~D.; Kapralov,~P.; Knyazev,~G.; Sekatskii,~S.; Dietler,~G.;
  Nur-E-Alam,~M.; Vasiliev,~M.; Alameh,~K.; Belotelov,~V. High-Q surface modes
  in photonic crystal/iron garnet film heterostructures for sensor
  applications. \emph{JETP letters} \textbf{2016}, \emph{104}, 679--684\relax
\mciteBstWouldAddEndPuncttrue
\mciteSetBstMidEndSepPunct{\mcitedefaultmidpunct}
{\mcitedefaultendpunct}{\mcitedefaultseppunct}\relax
\EndOfBibitem
\bibitem[Regatos \latin{et~al.}(2011)Regatos, Sep{u}lveda, Fari{n}a,
  Carrascosa, and Lechuga]{regatos2011suitable}
Regatos,~D.; Sep{u}lveda,~B.; Fari{n}a,~D.; Carrascosa,~L.~G.; Lechuga,~L.~M.
  Suitable combination of noble/ferromagnetic metal multilayers for enhanced
  magneto-plasmonic biosensing. \emph{Optics express} \textbf{2011}, \emph{19},
  8336--8346\relax
\mciteBstWouldAddEndPuncttrue
\mciteSetBstMidEndSepPunct{\mcitedefaultmidpunct}
{\mcitedefaultendpunct}{\mcitedefaultseppunct}\relax
\EndOfBibitem
\bibitem[Inoue \latin{et~al.}(2011)Inoue, Baryshev, Takagi, Lim, Hatafuku,
  Noda, and Togo]{inoue2011investigating}
Inoue,~M.; Baryshev,~A.; Takagi,~H.; Lim,~P.~B.; Hatafuku,~K.; Noda,~J.;
  Togo,~K. Investigating the use of magnonic crystals as extremely sensitive
  magnetic field sensors at room temperature. \emph{Applied Physics Letters}
  \textbf{2011}, \emph{98}, 132511\relax
\mciteBstWouldAddEndPuncttrue
\mciteSetBstMidEndSepPunct{\mcitedefaultmidpunct}
{\mcitedefaultendpunct}{\mcitedefaultseppunct}\relax
\EndOfBibitem
\bibitem[Baryshev \latin{et~al.}(2013)Baryshev, Merzlikin, and
  Inoue]{baryshev2013efficiency}
Baryshev,~A.; Merzlikin,~A.; Inoue,~M. Efficiency of optical sensing by a
  plasmonic photonic-crystal slab. \emph{Journal of Physics D: Applied Physics}
  \textbf{2013}, \emph{46}, 125107\relax
\mciteBstWouldAddEndPuncttrue
\mciteSetBstMidEndSepPunct{\mcitedefaultmidpunct}
{\mcitedefaultendpunct}{\mcitedefaultseppunct}\relax
\EndOfBibitem
\bibitem[Qin \latin{et~al.}(2017)Qin, Zhang, Liang, Liu, Wang, Kang, Lu, Zhang,
  Zhou, Wang, \latin{et~al.} others]{qin2017ultrahigh}
Qin,~J.; Zhang,~Y.; Liang,~X.; Liu,~C.; Wang,~C.; Kang,~T.; Lu,~H.; Zhang,~L.;
  Zhou,~P.; Wang,~X., \latin{et~al.}  Ultrahigh figure-of-merit in
  metal--insulator--metal magnetoplasmonic sensors using low loss
  magneto-optical oxide thin films. \emph{ACS Photonics} \textbf{2017},
  \emph{4}, 1403--1412\relax
\mciteBstWouldAddEndPuncttrue
\mciteSetBstMidEndSepPunct{\mcitedefaultmidpunct}
{\mcitedefaultendpunct}{\mcitedefaultseppunct}\relax
\EndOfBibitem
\bibitem[Ignatyeva \latin{et~al.}(2021)Ignatyeva, Knyazev, Kalish, Chernov, and
  Belotelov]{ignatyeva2021vector}
Ignatyeva,~D.~O.; Knyazev,~G.~A.; Kalish,~A.~N.; Chernov,~A.~I.;
  Belotelov,~V.~I. Vector magneto-optical magnetometer based on resonant
  all-dielectric gratings with highly anisotropic iron garnet films.
  \emph{Journal of Physics D: Applied Physics} \textbf{2021}, \emph{54},
  295001\relax
\mciteBstWouldAddEndPuncttrue
\mciteSetBstMidEndSepPunct{\mcitedefaultmidpunct}
{\mcitedefaultendpunct}{\mcitedefaultseppunct}\relax
\EndOfBibitem
\bibitem[Knyazev \latin{et~al.}(2018)Knyazev, Kapralov, Gusev, Kalish,
  Vetoshko, Dagesyan, Shaposhnikov, Prokopov, Berzhansky, Zvezdin,
  \latin{et~al.} others]{knyazev2018magnetoplasmonic}
Knyazev,~G.~A.; Kapralov,~P.~O.; Gusev,~N.~A.; Kalish,~A.~N.; Vetoshko,~P.~M.;
  Dagesyan,~S.~A.; Shaposhnikov,~A.~N.; Prokopov,~A.~R.; Berzhansky,~V.~N.;
  Zvezdin,~A.~K., \latin{et~al.}  Magnetoplasmonic crystals for highly
  sensitive magnetometry. \emph{ACS Photonics} \textbf{2018}, \emph{5},
  4951--4959\relax
\mciteBstWouldAddEndPuncttrue
\mciteSetBstMidEndSepPunct{\mcitedefaultmidpunct}
{\mcitedefaultendpunct}{\mcitedefaultseppunct}\relax
\EndOfBibitem
\bibitem[Postava \latin{et~al.}(2004)Postava, Pi{\v{s}}tora, and
  Yamaguchi]{postava2004magneto}
Postava,~K.; Pi{\v{s}}tora,~J.; Yamaguchi,~T. Magneto-optic vector magnetometry
  for sensor applications. \emph{Sensors and Actuators A: Physical}
  \textbf{2004}, \emph{110}, 242--246\relax
\mciteBstWouldAddEndPuncttrue
\mciteSetBstMidEndSepPunct{\mcitedefaultmidpunct}
{\mcitedefaultendpunct}{\mcitedefaultseppunct}\relax
\EndOfBibitem
\bibitem[Kuschel \latin{et~al.}(2011)Kuschel, Bardenhagen, Wilkens, Schubert,
  Hamrle, Pi{\v{s}}tora, and Wollschl{\"a}ger]{kuschel2011vectorial}
Kuschel,~T.; Bardenhagen,~H.; Wilkens,~H.; Schubert,~R.; Hamrle,~J.;
  Pi{\v{s}}tora,~J.; Wollschl{\"a}ger,~J. Vectorial magnetometry using
  magnetooptic Kerr effect including first-and second-order contributions for
  thin ferromagnetic films. \emph{Journal of Physics D: Applied Physics}
  \textbf{2011}, \emph{44}, 265003\relax
\mciteBstWouldAddEndPuncttrue
\mciteSetBstMidEndSepPunct{\mcitedefaultmidpunct}
{\mcitedefaultendpunct}{\mcitedefaultseppunct}\relax
\EndOfBibitem
\bibitem[Mikhailova \latin{et~al.}(2018)Mikhailova, Berzhansky, Shaposhnikov,
  Karavainikov, Prokopov, Kharchenko, Lukienko, Miloslavskaya, and
  Kharchenko]{mikhailova2018optimization}
Mikhailova,~T.; Berzhansky,~V.; Shaposhnikov,~A.; Karavainikov,~A.;
  Prokopov,~A.; Kharchenko,~Y.~M.; Lukienko,~I.; Miloslavskaya,~O.;
  Kharchenko,~M. Optimization of one-dimensional photonic crystals with double
  layer magneto-active defect. \emph{Optical Materials} \textbf{2018},
  \emph{78}, 521--530\relax
\mciteBstWouldAddEndPuncttrue
\mciteSetBstMidEndSepPunct{\mcitedefaultmidpunct}
{\mcitedefaultendpunct}{\mcitedefaultseppunct}\relax
\EndOfBibitem
\bibitem[Dong \latin{et~al.}(2020)Dong, Liu, Fei, Fan, Li, and
  Fu]{dong2020enhancing}
Dong,~D.; Liu,~Y.; Fei,~Y.; Fan,~Y.; Li,~J.; Fu,~Y. Enhancing the Faraday
  rotation in the monolayer phosphorus base of magneto-photonic crystals.
  \emph{Optical Materials} \textbf{2020}, \emph{102}, 109809\relax
\mciteBstWouldAddEndPuncttrue
\mciteSetBstMidEndSepPunct{\mcitedefaultmidpunct}
{\mcitedefaultendpunct}{\mcitedefaultseppunct}\relax
\EndOfBibitem
\bibitem[Steel \latin{et~al.}(2000)Steel, Levy, and Osgood]{steel2000photonic}
Steel,~M.; Levy,~M.; Osgood,~R. Photonic bandgaps with defects and the
  enhancement of Faraday rotation. \emph{Journal of lightwave technology}
  \textbf{2000}, \emph{18}, 1297\relax
\mciteBstWouldAddEndPuncttrue
\mciteSetBstMidEndSepPunct{\mcitedefaultmidpunct}
{\mcitedefaultendpunct}{\mcitedefaultseppunct}\relax
\EndOfBibitem
\bibitem[Lyubchanskii \latin{et~al.}(2006)Lyubchanskii, Dadoenkova,
  Lyubchanskii, Shapovalov, Zabolotin, Lee, and
  Rasing]{lyubchanskii2006response}
Lyubchanskii,~I.; Dadoenkova,~N.; Lyubchanskii,~M.; Shapovalov,~E.;
  Zabolotin,~A.; Lee,~Y.; Rasing,~T. Response of two-defect magnetic photonic
  crystals to oblique incidence of light: Effect of defect layer variation.
  2006\relax
\mciteBstWouldAddEndPuncttrue
\mciteSetBstMidEndSepPunct{\mcitedefaultmidpunct}
{\mcitedefaultendpunct}{\mcitedefaultseppunct}\relax
\EndOfBibitem
\bibitem[Grishin and Khartsev(2019)Grishin, and
  Khartsev]{grishin2019waveguiding}
Grishin,~A.~M.; Khartsev,~S. Waveguiding in all-garnet heteroepitaxial
  magneto-optical photonic crystals. \emph{JETP Letters} \textbf{2019},
  \emph{109}, 83--86\relax
\mciteBstWouldAddEndPuncttrue
\mciteSetBstMidEndSepPunct{\mcitedefaultmidpunct}
{\mcitedefaultendpunct}{\mcitedefaultseppunct}\relax
\EndOfBibitem
\bibitem[Inoue \latin{et~al.}(2006)Inoue, Fujikawa, Baryshev, Khanikaev, Lim,
  Uchida, Aktsipetrov, Fedyanin, Murzina, and Granovsky]{Inoue_2006}
Inoue,~M.; Fujikawa,~R.; Baryshev,~A.; Khanikaev,~A.; Lim,~P.~B.; Uchida,~H.;
  Aktsipetrov,~O.; Fedyanin,~A.; Murzina,~T.; Granovsky,~A. Magnetophotonic
  crystals. \emph{Journal of Physics D: Applied Physics} \textbf{2006},
  \emph{39}, R151--R161\relax
\mciteBstWouldAddEndPuncttrue
\mciteSetBstMidEndSepPunct{\mcitedefaultmidpunct}
{\mcitedefaultendpunct}{\mcitedefaultseppunct}\relax
\EndOfBibitem
\bibitem[Ignatyeva and Belotelov(2020)Ignatyeva, and
  Belotelov]{ignatyeva2020bound}
Ignatyeva,~D.; Belotelov,~V. Bound states in the continuum enable modulation of
  light intensity in the Faraday configuration. \emph{Optics Letters}
  \textbf{2020}, \emph{45}, 6422--6425\relax
\mciteBstWouldAddEndPuncttrue
\mciteSetBstMidEndSepPunct{\mcitedefaultmidpunct}
{\mcitedefaultendpunct}{\mcitedefaultseppunct}\relax
\EndOfBibitem
\bibitem[Dadoenkova \latin{et~al.}(2017)Dadoenkova, Dadoenkova, Lyubchanskii,
  K{\l}os, and Krawczyk]{dadoenkova2017faraday}
Dadoenkova,~Y.~S.; Dadoenkova,~N.~N.; Lyubchanskii,~I.~L.; K{\l}os,~J.~W.;
  Krawczyk,~M. Faraday effect in bi-periodic photonic-magnonic crystals.
  \emph{IEEE Transactions on Magnetics} \textbf{2017}, \emph{53}, 1--5\relax
\mciteBstWouldAddEndPuncttrue
\mciteSetBstMidEndSepPunct{\mcitedefaultmidpunct}
{\mcitedefaultendpunct}{\mcitedefaultseppunct}\relax
\EndOfBibitem
\bibitem[Dyakov \latin{et~al.}(2019)Dyakov, Fradkin, Gippius, Klompmaker,
  Spitzer, Yalcin, Akimov, Bayer, Yavsin, Pavlov, \latin{et~al.}
  others]{dyakov2019wide}
Dyakov,~S.; Fradkin,~I.; Gippius,~N.; Klompmaker,~L.; Spitzer,~F.; Yalcin,~E.;
  Akimov,~I.; Bayer,~M.; Yavsin,~D.; Pavlov,~S., \latin{et~al.}  Wide-band
  enhancement of the transverse magneto-optical Kerr effect in magnetite-based
  plasmonic crystals. \emph{Physical Review B} \textbf{2019}, \emph{100},
  214411\relax
\mciteBstWouldAddEndPuncttrue
\mciteSetBstMidEndSepPunct{\mcitedefaultmidpunct}
{\mcitedefaultendpunct}{\mcitedefaultseppunct}\relax
\EndOfBibitem
\bibitem[Pavlov \latin{et~al.}(2008)Pavlov, Usachev, Pisarev, Kurdyukov,
  Kaplan, Kimel, Kirilyuk, and Rasing]{pavlov2008enhancement}
Pavlov,~V.; Usachev,~P.; Pisarev,~R.; Kurdyukov,~D.; Kaplan,~S.; Kimel,~A.;
  Kirilyuk,~A.; Rasing,~T. Enhancement of optical and magneto-optical effects
  in three-dimensional opal/Fe 3 O 4 magnetic photonic crystals. \emph{Applied
  physics letters} \textbf{2008}, \emph{93}, 072502\relax
\mciteBstWouldAddEndPuncttrue
\mciteSetBstMidEndSepPunct{\mcitedefaultmidpunct}
{\mcitedefaultendpunct}{\mcitedefaultseppunct}\relax
\EndOfBibitem
\bibitem[Maccaferri \latin{et~al.}(2020)Maccaferri, Zubritskaya, Razdolski,
  Chioar, Belotelov, Kapaklis, Oppeneer, and Dmitriev]{maccaferri2020nanoscale}
Maccaferri,~N.; Zubritskaya,~I.; Razdolski,~I.; Chioar,~I.-A.; Belotelov,~V.;
  Kapaklis,~V.; Oppeneer,~P.~M.; Dmitriev,~A. Nanoscale magnetophotonics.
  \emph{Journal of Applied Physics} \textbf{2020}, \emph{127}, 080903\relax
\mciteBstWouldAddEndPuncttrue
\mciteSetBstMidEndSepPunct{\mcitedefaultmidpunct}
{\mcitedefaultendpunct}{\mcitedefaultseppunct}\relax
\EndOfBibitem
\bibitem[Caballero \latin{et~al.}(2015)Caballero, Garcia-Martin, and
  Cuevas]{caballero2015faraday}
Caballero,~B.; Garcia-Martin,~A.; Cuevas,~J. Faraday effect in hybrid
  magneto-plasmonic photonic crystals. \emph{Optics express} \textbf{2015},
  \emph{23}, 22238--22249\relax
\mciteBstWouldAddEndPuncttrue
\mciteSetBstMidEndSepPunct{\mcitedefaultmidpunct}
{\mcitedefaultendpunct}{\mcitedefaultseppunct}\relax
\EndOfBibitem
\bibitem[Hamidi \latin{et~al.}(2015)Hamidi, Razavinia, and
  Tehranchi]{hamidi2015enhanced}
Hamidi,~S.; Razavinia,~M.; Tehranchi,~M. Enhanced optically induced
  magnetization due to inverse Faraday effect in plasmonic nanostructures.
  \emph{Optics Communications} \textbf{2015}, \emph{338}, 240--245\relax
\mciteBstWouldAddEndPuncttrue
\mciteSetBstMidEndSepPunct{\mcitedefaultmidpunct}
{\mcitedefaultendpunct}{\mcitedefaultseppunct}\relax
\EndOfBibitem
\bibitem[Kuzmin \latin{et~al.}(2016)Kuzmin, Bychkov, Shavrov, and
  Temnov]{kuzmin2016giant}
Kuzmin,~D.~A.; Bychkov,~I.~V.; Shavrov,~V.~G.; Temnov,~V.~V. Giant Faraday
  rotation of high-order plasmonic modes in graphene-covered nanowires.
  \emph{Nano letters} \textbf{2016}, \emph{16}, 4391--4395\relax
\mciteBstWouldAddEndPuncttrue
\mciteSetBstMidEndSepPunct{\mcitedefaultmidpunct}
{\mcitedefaultendpunct}{\mcitedefaultseppunct}\relax
\EndOfBibitem
\bibitem[Uchida \latin{et~al.}(2009)Uchida, Masuda, Fujikawa, Baryshev, and
  Inoue]{uchida2009large}
Uchida,~H.; Masuda,~Y.; Fujikawa,~R.; Baryshev,~A.; Inoue,~M. Large enhancement
  of Faraday rotation by localized surface plasmon resonance in Au
  nanoparticles embedded in Bi: YIG film. \emph{Journal of Magnetism and
  Magnetic Materials} \textbf{2009}, \emph{321}, 843--845\relax
\mciteBstWouldAddEndPuncttrue
\mciteSetBstMidEndSepPunct{\mcitedefaultmidpunct}
{\mcitedefaultendpunct}{\mcitedefaultseppunct}\relax
\EndOfBibitem
\bibitem[Chekhov \latin{et~al.}(2018)Chekhov, Naydenov, Smirnova, Ketsko,
  Stognij, and Murzina]{chekhov2018magnetoplasmonic}
Chekhov,~A.; Naydenov,~P.; Smirnova,~M.; Ketsko,~V.; Stognij,~A.; Murzina,~T.
  Magnetoplasmonic crystal waveguide. \emph{Optics express} \textbf{2018},
  \emph{26}, 21086--21091\relax
\mciteBstWouldAddEndPuncttrue
\mciteSetBstMidEndSepPunct{\mcitedefaultmidpunct}
{\mcitedefaultendpunct}{\mcitedefaultseppunct}\relax
\EndOfBibitem
\bibitem[Pomozov \latin{et~al.}(2020)Pomozov, Chekhov, Rodionov, Baburin,
  Lotkov, Temiryazeva, Afanasyev, Baryshev, and Murzina]{pomozov2020two}
Pomozov,~A.; Chekhov,~A.; Rodionov,~I.; Baburin,~A.; Lotkov,~E.;
  Temiryazeva,~M.; Afanasyev,~K.; Baryshev,~A.; Murzina,~T. Two-dimensional
  high-quality Ag/Py magnetoplasmonic crystals. \emph{Applied Physics Letters}
  \textbf{2020}, \emph{116}, 013106\relax
\mciteBstWouldAddEndPuncttrue
\mciteSetBstMidEndSepPunct{\mcitedefaultmidpunct}
{\mcitedefaultendpunct}{\mcitedefaultseppunct}\relax
\EndOfBibitem
\bibitem[Belotelov and Zvezdin(2006)Belotelov, and
  Zvezdin]{belotelov2006magnetooptics}
Belotelov,~V.; Zvezdin,~A. Magnetooptics and extraordinary transmission of the
  perforated metallic films magnetized in polar geometry. \emph{Journal of
  Magnetism and Magnetic Materials} \textbf{2006}, \emph{300}, e260--e263\relax
\mciteBstWouldAddEndPuncttrue
\mciteSetBstMidEndSepPunct{\mcitedefaultmidpunct}
{\mcitedefaultendpunct}{\mcitedefaultseppunct}\relax
\EndOfBibitem
\bibitem[Borovkova \latin{et~al.}(2018)Borovkova, Hashim, Kozhaev, Dagesyan,
  Chakravarty, Levy, and Belotelov]{borovkova2018tmoke}
Borovkova,~O.; Hashim,~H.; Kozhaev,~M.; Dagesyan,~S.; Chakravarty,~A.;
  Levy,~M.; Belotelov,~V. TMOKE as efficient tool for the magneto-optic
  analysis of ultra-thin magnetic films. \emph{Applied Physics Letters}
  \textbf{2018}, \emph{112}, 063101\relax
\mciteBstWouldAddEndPuncttrue
\mciteSetBstMidEndSepPunct{\mcitedefaultmidpunct}
{\mcitedefaultendpunct}{\mcitedefaultseppunct}\relax
\EndOfBibitem
\bibitem[Razdolski \latin{et~al.}(2013)Razdolski, Gheorghe, Melander,
  Hj{\"o}rvarsson, Patoka, Kimel, Kirilyuk, Papaioannou, and
  Rasing]{razdolski2013nonlocal}
Razdolski,~I.; Gheorghe,~D.; Melander,~E.; Hj{\"o}rvarsson,~B.; Patoka,~P.;
  Kimel,~A.; Kirilyuk,~A.; Papaioannou,~E.~T.; Rasing,~T. Nonlocal nonlinear
  magneto-optical response of a magnetoplasmonic crystal. \emph{Physical Review
  B} \textbf{2013}, \emph{88}, 075436\relax
\mciteBstWouldAddEndPuncttrue
\mciteSetBstMidEndSepPunct{\mcitedefaultmidpunct}
{\mcitedefaultendpunct}{\mcitedefaultseppunct}\relax
\EndOfBibitem
\bibitem[Kharratian \latin{et~al.}(2020)Kharratian, Urey, and
  Onbasli]{kharratian2020broadband}
Kharratian,~S.; Urey,~H.; Onbasli,~M.~C. Broadband Enhancement of Faraday
  Effect Using Magnetoplasmonic Metasurfaces. \emph{Plasmonics} \textbf{2020},
  1--11\relax
\mciteBstWouldAddEndPuncttrue
\mciteSetBstMidEndSepPunct{\mcitedefaultmidpunct}
{\mcitedefaultendpunct}{\mcitedefaultseppunct}\relax
\EndOfBibitem
\bibitem[Gabbani \latin{et~al.}(2021)Gabbani, Fantechi, Petrucci, Campo,
  de~Julian~Fernandez, Ghigna, Sorace, Bonanni, Gurioli, Sangregorio,
  \latin{et~al.} others]{gabbani2021dielectric}
Gabbani,~A.; Fantechi,~E.; Petrucci,~G.; Campo,~G.; de~Julian~Fernandez,~C.;
  Ghigna,~P.; Sorace,~L.; Bonanni,~V.; Gurioli,~M.; Sangregorio,~C.,
  \latin{et~al.}  Dielectric Effects in FeO x-Coated Au Nanoparticles Boost the
  Magnetoplasmonic Response: Implications for Active Plasmonic Devices.
  \emph{ACS Applied Nano Materials} \textbf{2021}, \relax
\mciteBstWouldAddEndPunctfalse
\mciteSetBstMidEndSepPunct{\mcitedefaultmidpunct}
{}{\mcitedefaultseppunct}\relax
\EndOfBibitem
\bibitem[Zubritskaya \latin{et~al.}(2018)Zubritskaya, Maccaferri,
  Inchausti~Ezeiza, Vavassori, and Dmitriev]{zubritskaya2018magnetic}
Zubritskaya,~I.; Maccaferri,~N.; Inchausti~Ezeiza,~X.; Vavassori,~P.;
  Dmitriev,~A. Magnetic control of the chiroptical plasmonic surfaces.
  \emph{Nano letters} \textbf{2018}, \emph{18}, 302--307\relax
\mciteBstWouldAddEndPuncttrue
\mciteSetBstMidEndSepPunct{\mcitedefaultmidpunct}
{\mcitedefaultendpunct}{\mcitedefaultseppunct}\relax
\EndOfBibitem
\bibitem[Pourjamal \latin{et~al.}(2019)Pourjamal, Kataja, Maccaferri,
  Vavassori, and van Dijken]{pourjamal2019tunable}
Pourjamal,~S.; Kataja,~M.; Maccaferri,~N.; Vavassori,~P.; van Dijken,~S.
  Tunable magnetoplasmonics in lattices of Ni/SiO 2/Au dimers. \emph{Scientific
  reports} \textbf{2019}, \emph{9}, 1--11\relax
\mciteBstWouldAddEndPuncttrue
\mciteSetBstMidEndSepPunct{\mcitedefaultmidpunct}
{\mcitedefaultendpunct}{\mcitedefaultseppunct}\relax
\EndOfBibitem
\bibitem[L{\'o}pez-Ortega \latin{et~al.}(2020)L{\'o}pez-Ortega, Zapata-Herrera,
  Maccaferri, Pancaldi, Garcia, Chuvilin, and Vavassori]{lopez2020enhanced}
L{\'o}pez-Ortega,~A.; Zapata-Herrera,~M.; Maccaferri,~N.; Pancaldi,~M.;
  Garcia,~M.; Chuvilin,~A.; Vavassori,~P. Enhanced magnetic modulation of light
  polarization exploiting hybridization with multipolar dark plasmons in
  magnetoplasmonic nanocavities. \emph{Light: Science \& Applications}
  \textbf{2020}, \emph{9}, 1--14\relax
\mciteBstWouldAddEndPuncttrue
\mciteSetBstMidEndSepPunct{\mcitedefaultmidpunct}
{\mcitedefaultendpunct}{\mcitedefaultseppunct}\relax
\EndOfBibitem
\bibitem[Kolmychek \latin{et~al.}(2018)Kolmychek, Pomozov, Leontiev, Napolskii,
  and Murzina]{kolmychek2018magneto}
Kolmychek,~I.; Pomozov,~A.; Leontiev,~A.; Napolskii,~K.; Murzina,~T.
  Magneto-optical effects in hyperbolic metamaterials. \emph{Optics letters}
  \textbf{2018}, \emph{43}, 3917--3920\relax
\mciteBstWouldAddEndPuncttrue
\mciteSetBstMidEndSepPunct{\mcitedefaultmidpunct}
{\mcitedefaultendpunct}{\mcitedefaultseppunct}\relax
\EndOfBibitem
\bibitem[Chin \latin{et~al.}(2013)Chin, Steinle, Wehlus, Dregely, Weiss,
  Belotelov, Stritzker, and Giessen]{chin2013nonreciprocal}
Chin,~J.~Y.; Steinle,~T.; Wehlus,~T.; Dregely,~D.; Weiss,~T.; Belotelov,~V.~I.;
  Stritzker,~B.; Giessen,~H. Nonreciprocal plasmonics enables giant enhancement
  of thin-film Faraday rotation. \emph{Nature communications} \textbf{2013},
  \emph{4}, 1--6\relax
\mciteBstWouldAddEndPuncttrue
\mciteSetBstMidEndSepPunct{\mcitedefaultmidpunct}
{\mcitedefaultendpunct}{\mcitedefaultseppunct}\relax
\EndOfBibitem
\bibitem[Khramova \latin{et~al.}(2019)Khramova, Ignatyeva, Kozhaev, Dagesyan,
  Berzhansky, Shaposhnikov, Tomilin, and Belotelov]{khramova2019resonances}
Khramova,~A.~E.; Ignatyeva,~D.~O.; Kozhaev,~M.~A.; Dagesyan,~S.~A.;
  Berzhansky,~V.~N.; Shaposhnikov,~A.~N.; Tomilin,~S.~V.; Belotelov,~V.~I.
  Resonances of the magneto-optical intensity effect mediated by interaction of
  different modes in a hybrid magnetoplasmonic heterostructure with gold
  nanoparticles. \emph{Optics express} \textbf{2019}, \emph{27},
  33170--33179\relax
\mciteBstWouldAddEndPuncttrue
\mciteSetBstMidEndSepPunct{\mcitedefaultmidpunct}
{\mcitedefaultendpunct}{\mcitedefaultseppunct}\relax
\EndOfBibitem
\bibitem[Royer \latin{et~al.}(2020)Royer, Varghese, Gamet, Neveu, Jourlin, and
  Jamon]{royer2020enhancement}
Royer,~F.; Varghese,~B.; Gamet,~E.; Neveu,~S.; Jourlin,~Y.; Jamon,~D.
  Enhancement of both Faraday and Kerr effects with an all-dielectric grating
  based on a magneto-optical nanocomposite material. \emph{ACS omega}
  \textbf{2020}, \emph{5}, 2886--2892\relax
\mciteBstWouldAddEndPuncttrue
\mciteSetBstMidEndSepPunct{\mcitedefaultmidpunct}
{\mcitedefaultendpunct}{\mcitedefaultseppunct}\relax
\EndOfBibitem
\bibitem[Bsawmaii \latin{et~al.}(2020)Bsawmaii, Gamet, Royer, Neveu, and
  Jamon]{bsawmaii2020longitudinal}
Bsawmaii,~L.; Gamet,~E.; Royer,~F.; Neveu,~S.; Jamon,~D. Longitudinal
  magneto-optical effect enhancement with high transmission through a 1D
  all-dielectric resonant guided mode grating. \emph{Optics express}
  \textbf{2020}, \emph{28}, 8436--8444\relax
\mciteBstWouldAddEndPuncttrue
\mciteSetBstMidEndSepPunct{\mcitedefaultmidpunct}
{\mcitedefaultendpunct}{\mcitedefaultseppunct}\relax
\EndOfBibitem
\bibitem[Voronov \latin{et~al.}(2020)Voronov, Karki, Ignatyeva, Kozhaev, Levy,
  and Belotelov]{voronov2020magneto}
Voronov,~A.~A.; Karki,~D.; Ignatyeva,~D.~O.; Kozhaev,~M.~A.; Levy,~M.;
  Belotelov,~V.~I. Magneto-optics of subwavelength all-dielectric gratings.
  \emph{Optics Express} \textbf{2020}, \emph{28}, 17988--17996\relax
\mciteBstWouldAddEndPuncttrue
\mciteSetBstMidEndSepPunct{\mcitedefaultmidpunct}
{\mcitedefaultendpunct}{\mcitedefaultseppunct}\relax
\EndOfBibitem
\bibitem[Chernov \latin{et~al.}(2020)Chernov, Kozhaev, Ignatyeva, Beginin,
  Sadovnikov, Voronov, Karki, Levy, and Belotelov]{chernov2020all}
Chernov,~A.~I.; Kozhaev,~M.~A.; Ignatyeva,~D.~O.; Beginin,~E.~N.;
  Sadovnikov,~A.~V.; Voronov,~A.~A.; Karki,~D.; Levy,~M.; Belotelov,~V.~I.
  All-dielectric nanophotonics enables tunable excitation of the exchange spin
  waves. \emph{Nano letters} \textbf{2020}, \emph{20}, 5259--5266\relax
\mciteBstWouldAddEndPuncttrue
\mciteSetBstMidEndSepPunct{\mcitedefaultmidpunct}
{\mcitedefaultendpunct}{\mcitedefaultseppunct}\relax
\EndOfBibitem
\bibitem[Ignatyeva \latin{et~al.}(2020)Ignatyeva, Karki, Voronov, Kozhaev,
  Krichevsky, Chernov, Levy, and Belotelov]{ignatyeva2020all}
Ignatyeva,~D.~O.; Karki,~D.; Voronov,~A.~A.; Kozhaev,~M.~A.; Krichevsky,~D.~M.;
  Chernov,~A.~I.; Levy,~M.; Belotelov,~V.~I. All-dielectric magnetic
  metasurface for advanced light control in dual polarizations combined with
  high-Q resonances. \emph{Nature communications} \textbf{2020}, \emph{11},
  1--8\relax
\mciteBstWouldAddEndPuncttrue
\mciteSetBstMidEndSepPunct{\mcitedefaultmidpunct}
{\mcitedefaultendpunct}{\mcitedefaultseppunct}\relax
\EndOfBibitem
\bibitem[Adams(1981)]{adams1981introduction}
Adams,~M.~J. An introduction to optical waveguides. \textbf{1981}, \relax
\mciteBstWouldAddEndPunctfalse
\mciteSetBstMidEndSepPunct{\mcitedefaultmidpunct}
{}{\mcitedefaultseppunct}\relax
\EndOfBibitem
\bibitem[Snyder and Love(1986)Snyder, and Love]{snyder1986optical}
Snyder,~A.~W.; Love,~J.~D. Optical waveguide theory. \emph{J. Opt. Soc. Am. A}
  \textbf{1986}, \emph{3}, 378\relax
\mciteBstWouldAddEndPuncttrue
\mciteSetBstMidEndSepPunct{\mcitedefaultmidpunct}
{\mcitedefaultendpunct}{\mcitedefaultseppunct}\relax
\EndOfBibitem
\bibitem[Gloge(1971)]{gloge1971weakly}
Gloge,~D. Weakly guiding fibers. \emph{Applied optics} \textbf{1971},
  \emph{10}, 2252--2258\relax
\mciteBstWouldAddEndPuncttrue
\mciteSetBstMidEndSepPunct{\mcitedefaultmidpunct}
{\mcitedefaultendpunct}{\mcitedefaultseppunct}\relax
\EndOfBibitem
\bibitem[Rajeeva \latin{et~al.}(2018)Rajeeva, Lin, and
  Zheng]{rajeeva2018design}
Rajeeva,~B.~B.; Lin,~L.; Zheng,~Y. Design and applications of lattice plasmon
  resonances. \emph{Nano Research} \textbf{2018}, \emph{11}, 4423--4440\relax
\mciteBstWouldAddEndPuncttrue
\mciteSetBstMidEndSepPunct{\mcitedefaultmidpunct}
{\mcitedefaultendpunct}{\mcitedefaultseppunct}\relax
\EndOfBibitem
\bibitem[Kalish \latin{et~al.}(2014)Kalish, Ignatyeva, Belotelov, Kreilkamp,
  Akimov, Gopal, Bayer, and Sukhorukov]{kalish2014transformation}
Kalish,~A.~N.; Ignatyeva,~D.~O.; Belotelov,~V.~I.; Kreilkamp,~L.~E.;
  Akimov,~I.~A.; Gopal,~A.~V.; Bayer,~M.; Sukhorukov,~A.~P. Transformation of
  mode polarization in gyrotropic plasmonic waveguides. \emph{Laser Physics}
  \textbf{2014}, \emph{24}, 094006\relax
\mciteBstWouldAddEndPuncttrue
\mciteSetBstMidEndSepPunct{\mcitedefaultmidpunct}
{\mcitedefaultendpunct}{\mcitedefaultseppunct}\relax
\EndOfBibitem
\bibitem[Ignatyeva \latin{et~al.}(2012)Ignatyeva, Kalish, Levkina, and
  Sukhorukov]{ignatyeva2012surface}
Ignatyeva,~D.~O.; Kalish,~A.~N.; Levkina,~G.~Y.; Sukhorukov,~A.~P. Surface
  plasmon polaritons at gyrotropic interfaces. \emph{Physical Review A}
  \textbf{2012}, \emph{85}, 043804\relax
\mciteBstWouldAddEndPuncttrue
\mciteSetBstMidEndSepPunct{\mcitedefaultmidpunct}
{\mcitedefaultendpunct}{\mcitedefaultseppunct}\relax
\EndOfBibitem
\bibitem[Prokhorov \latin{et~al.}(1984)Prokhorov, Smolenskii, and
  Ageev]{prokhorov1984optical}
Prokhorov,~A.; Smolenskii,~G.; Ageev,~A. Optical phenomena in thin-film
  magnetic waveguides and their technical application. \emph{Soviet Physics
  Uspekhi} \textbf{1984}, \emph{27}, 339\relax
\mciteBstWouldAddEndPuncttrue
\mciteSetBstMidEndSepPunct{\mcitedefaultmidpunct}
{\mcitedefaultendpunct}{\mcitedefaultseppunct}\relax
\EndOfBibitem
\end{mcitethebibliography}

\end{document}